\documentclass[12pt]{iopart}

\usepackage{amsthm}
\usepackage{graphicx}

\theoremstyle{plain}
\newtheorem{theorem}{Theorem}

\theoremstyle{definition}
\newtheorem{defn}{Definition}
\newtheorem{rem}{Remark}

\newcommand{\U}{{\bf U}}
\newcommand{\E}{{\sf E}}
\newcommand{\M}{{\bf M}}
\newcommand{\B}{{\bf B}}
\newcommand{\A}{{\bf A}}
\newcommand{\brf}[1]{\left\{#1\right\}}
\newcommand{\brr}[1]{\left(#1\right)}

\begin{document}

\title[Spectral gap of doubly stochastic matrices generated from CUE]
{Spectral gap of doubly stochastic matrices generated from
  equidistributed unitary matrices}
\author{G Berkolaiko}
\address{Department of Complex Systems,
Weizmann Institute of Science,
Rehovot 76100, Israel}
\ead{Gregory.Berkolaiko@weizmann.ac.il}
\date{April 2, 2001}

\begin{abstract}
  To a unitary matrix $\U$ we associate a doubly stochastic matrix
  $\M$ by taking the modulus squared of each element of $\U$.  To
  study the connection between onset of quantum chaos on graphs and
  ergodicity of the underlying Markov chain, specified by $\M$, we
  study the limiting distribution of the spectral gap of $\M$ when
  $\U$ is taken from the Circular Unitary Ensemble and the dimension
  $N$ of $\U$ is taken to infinity.  We prove that the limiting
  distribution is degenerate: the gap tends to its maximal value 1.
  The shape of the gap distribution for finite $N$ is also discussed.
\end{abstract}

\submitto{\JPA}

\maketitle


\section{Introduction}

Recently it was proposed by Kottos and Smilansky \cite{KotS1, KotS2} that
quantum graphs can serve as a good ``toy'' model for quantum chaos.
Numerical simulations confirmed that for some classes of graphs,
including the fully connected graphs, the Bohigas-Giannoni-Schmit
Conjecture \cite{BohGS} holds true.  By this we mean that various
statistical functions of the spectrum of the graphs converge to the
corresponding functions obtained in Random Matrix Theory (RMT)
\cite{Meh} in the limit as the number of vertices is taken to infinity.

On the graphs, the quantum evolution operator assumes the form of a
finite unitary matrix $\U$.  In \cite{KotS1, KotS2} it was
argued that the classical counterpart of the quantum system is the
Markov chain on the directed bonds of the graph with the matrix $\M$
of the transition probabilities given by
\begin{equation}
  \label{eq:trans_prob}
  M_{b,b'} = |U_{b,b'}|^2.
\end{equation}
It was conjectured that there is a link between ergodicity of the
Markov chain and the convergence of the quantum statistics to RMT
results.

However, in \cite{BerK} it was shown that the statistics for a special
family of graphs, the quantum star graphs, converge to a non-RMT limit
as the size of the graphs increases.  However ``classically'' the star
graphs have nice properties: the corresponding Markov chain possesses
a unique attracting equilibrium distribution which corresponds to the
eigenvalue 1 of $\M$.  Thus ergodicity of each graph does {\it not}
immediately imply RMT-like behaviour.
 
It is well-known that a good quantitative measure of ergodicity of a
Markov chain is its spectral gap.  Given the spectrum $\{1, \lambda_2,
\lambda_3, \ldots \lambda_{N}\}$ of $\M$, the spectral gap is defined
as
\begin{equation}
  \label{eq:gap_defn}
  g = 1 - \max_{i=2,\ldots N} |\lambda_i|.
\end{equation}
The spectral gap indicates the speed of convergence of any initial
distribution to the equilibrium one: the greater $g$ is, the faster is
the convergence.  Tanner \cite{Tan, Tan_p} suggested that if for a
sequence of graphs the spectral gap of the corresponding Markov chains
is uniformly bounded away from zero, then the spectral statistics will
converge to their RMT form.  This seems to be a good indicator of the
conformance to RMT: while for each star graph the gap is non-zero, it
decreases to zero as we increase the size of the graph.  And in the
numerically verified cases of RMT-like behaviour the gap stays bounded
away from zero.

%

To investigate the connection between random unitary matrices and the
ergodicity of their ``classical'' analogues, we ask the following
question: given a large random unitary matrix, find the probability
distribution of the spectral gap of the corresponding Markov chain.

\section{Preliminaries}

Since the most natural ensemble of the unitary matrices is the
Circular Unitary Ensemble (CUE), we shall restrict our attention to
it, although the generalisation of the problem to other ensembles is
straightforward.

\begin{defn}
  CUE$(N)$ is defined as the ensemble of all unitary $N\times N$
  matrices endowed with the probability measure that is invariant
  under every automorphism
  \begin{equation}
    \label{eq:autom}
    \U \mapsto {\bf VUW },
  \end{equation}
  where ${\bf V}$ and ${\bf W}$ are any two $N\times N$ unitary
  matrices.
\end{defn}

\begin{defn}
  An entrywise nonnegative $N\times N$ matrix $\M$ is called {\sl
    doubly stochastic} if 
  \begin{equation}
    \sum_{i=1}^N M_{i,j} = 1\quad  \forall j
    \qquad\mbox{ and }\qquad 
    \sum_{j=1}^N M_{i,j} = 1\quad  \forall i.
  \end{equation}  
  The set of all such $N\times N$ matrices we denote by DS$(N)$.
\end{defn}

If $\U$ is unitary then the matrix $\M$ defined by
(\ref{eq:trans_prob}) is doubly stochastic.  For a Markov chain it
means that the uniform measure is invariant, that is $(1/N,\ldots,1/N)$
is a left eigenvector of $\M$ with the eigenvalue 1.

\begin{rem}
  It is easy to check that if $\A,\B\in \mbox{DS}(N)$ then $\A\B \in
  \mbox{DS}(N)$, thus DS$(N)$ forms a semigroup.
\end{rem}

Now we define the function $S: \mbox{CUE}(N) \to \mbox{DS}(N)$ to map
$\U$ to $\M$ according to~(\ref{eq:trans_prob}).  It is clear that $S$
induces a probability measure on the semigroup DS$(N)$.  Thus our
question is 
\begin{quote}
  what is the probability distribution of the spectral
  gap $g$ if the matrices $M$ are selected randomly with the
  probability induced by the correspondence $S$.
\end{quote}
We denote the gap probability density function by $f_N(g)$ and the
corresponding cumulative distribution by $F_N(g)=\int_0^g f_N(g')dg'$.
For convenience we also denote by $\lambda_2(\M)$ the second largest
(by modulus) eigenvalue of the doubly stochastic matrix $\M$,
\begin{equation}
  \label{eq:order}
  |\lambda_2(\M)| = \max_{i=2,\ldots N} |\lambda_i|
\end{equation}
and, as a consequence, $g=1-|\lambda_2(\M)|$.

\section{Numerics and a conjecture}
First of all we are going to present the results of some numerical
simulations.  The algorithm we used is simple: random matrices from
CUE$(N)$ are generated using the Hurwitz parametrisation \cite{Hur}
(see also \cite{PozZK}), the spectrum of the corresponding doubly
stochastic matrix $M$ is computed and the gap is calculated according
to the definition~(\ref{eq:gap_defn}).

The cumulative distribution functions obtained for $N=2$, $3$, $5$,
$10$, $20$, $40$, $80$ for sample sizes of $10^5$ are shown on
\Fref{fig:hist1}.  From the plot we can see that the typical gap tends
to 1 as we increase $N$.  Thus it is natural to conjecture that the
gap probability density function $f_N(g)$ converges to $\delta(g-1)$
in the sense of distributions as $N\to\infty$ or, in other words,
\begin{equation}
  \label{eq:prob_decay}
  \Pr\brf{0\leq g \leq a} \to 0 \qquad \mbox{as }N\to\infty \qquad
  \forall a<1.
\end{equation}

\begin{figure}[b]
  \includegraphics[scale=0.6]{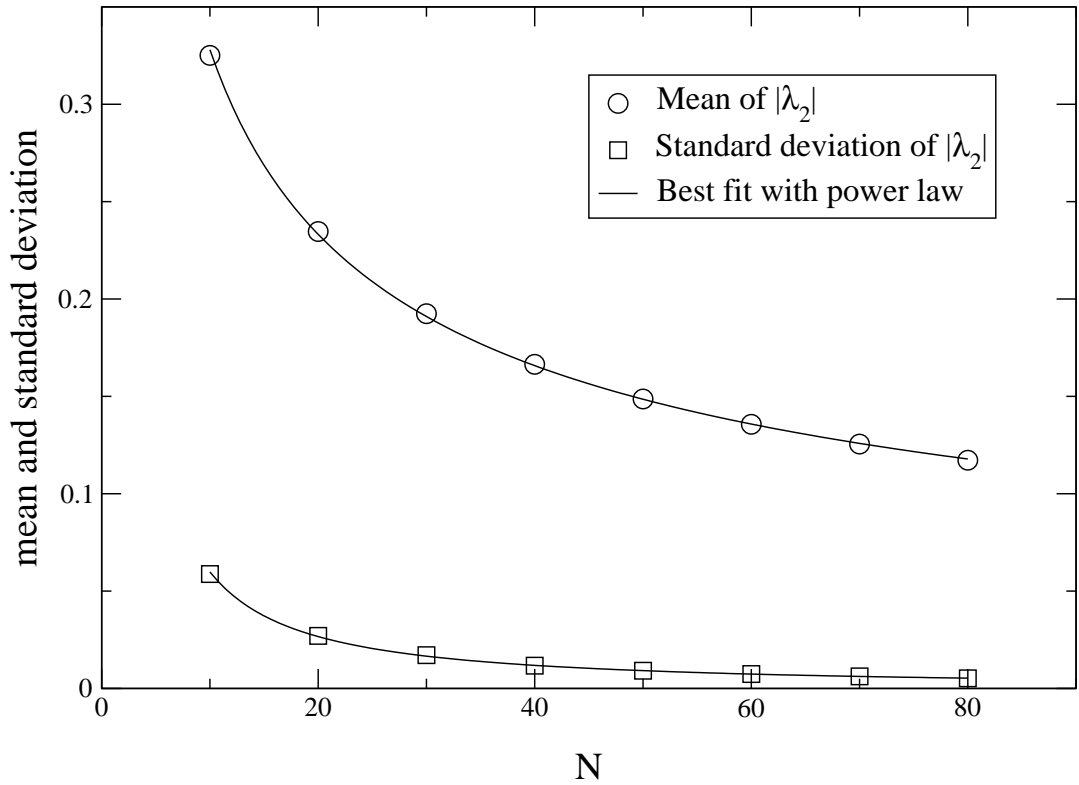}
  \vskip 0.5cm
  \caption{Estimations of the mean (circles) and the standard
    deviation (squares) of the modulus of the second largest
    eigenvalue $\lambda_2(\M)$ as functions of $N$.  The error bars
    (based on the 95\% confidence interval) are too small to be
    indicated: they are of order 1.5\% for the deviation and less
    than 0.2\% for the mean.  The lines correspond to the best fit
    with the power law.  The exponents are $-0.49$ for the mean and
    $-1.167$ for the deviation.}
  \label{fig:mean_and_std}
\end{figure}

Further evidence in support of this guess is presented by
\Fref{fig:mean_and_std}, where the mean $\E|\lambda_2|$ 
and the standard deviation $\sigma$ of the second largest eigenvalue
are plotted as functions of $N$.  A 
good fit to the mean is given by $\E|\lambda_2|\propto
1/\sqrt{N}$ (which is 
supported by heuristic observations presented in
Section~\ref{sec:verify}).  Note that since $0\leq |\lambda_2|\leq1$
and
\begin{equation}
  \label{eq:obv_bound}
  \E|\lambda_2| \geq a \Pr\brf{a\leq |\lambda_2| \leq 1} + 0
  \Pr\brf{0\leq |\lambda_2| \leq a} = a\Pr\brf{0\leq g \leq a},
\end{equation}
to verify (\ref{eq:prob_decay}) it is enough to show that
$\E|\lambda_2|\to0$ as $N\to\infty$.

\section{Verification of the
  conjecture}
\label{sec:verify}

\subsection{$N=2$ case}

\begin{figure}[c]
  \includegraphics[scale=0.6]{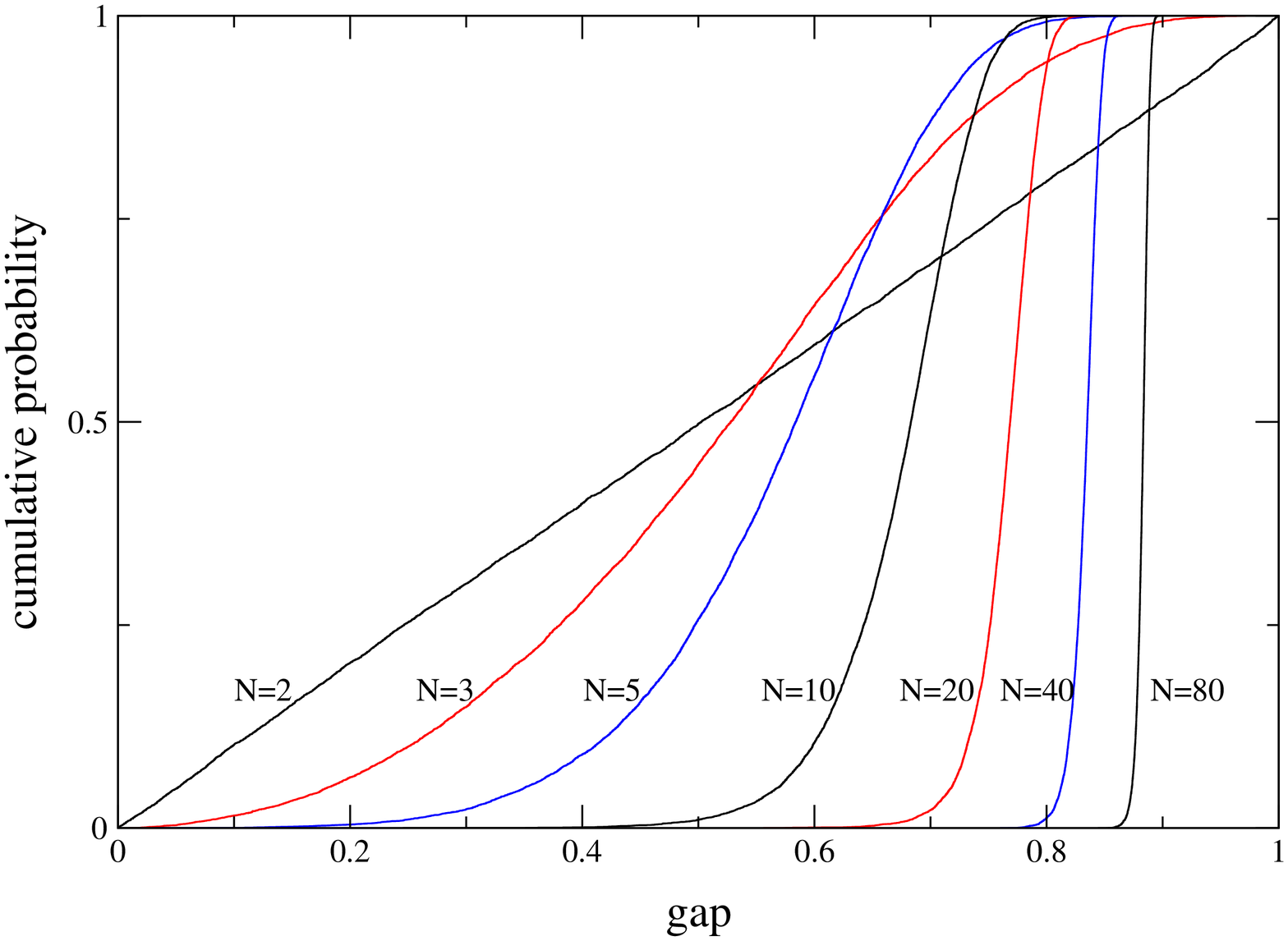}
  \vskip0.5cm
  \caption{Cumulative distribution function of the gaps of the doubly
    stochastic matrices obtained from the unitary matrices from
    CUE$(N)$ for different values of $N$. } 
  \label{fig:hist1}
\end{figure}

In the case of $N=2$ it is easy to calculate the gap probability
explicitly.  For completeness we include the derivation although it is
obvious from \Fref{fig:hist1} that the distribution for $N=2$ is not
typical.  A $2\times2$ unitary matrix can be parametrized as
\begin{equation}
  \label{eq:2dim_unit}
  \U = e^{i\alpha}
  \left(
    \begin{array}{cc}
      e^{i\psi} \cos\phi  & e^{i\chi} \sin\phi \\
      -e^{i\chi} \sin\phi & e^{-i\psi} \cos\phi 
    \end{array}
  \right),
\end{equation}
with
\begin{equation}
  \label{eq:intervals}
  0\leq \alpha\leq 2\pi, \quad 
  0\leq \chi  \leq 2\pi, \quad
  0\leq \psi  \leq 2\pi, \quad
  0\leq \phi  \leq \pi/2.
\end{equation}
The Haar (uniform) measure on the CUE(2) can be expressed in terms of
the parameters $\alpha$, $\chi$, $\psi$ and $\phi$ as $(2\pi)^{-3}d\alpha
\,d\chi\,d\psi\,d\sin^2\!\phi$.  The corresponding doubly stochastic
matrix is 
\begin{equation}
  \label{eq:2dim_bis}
  \M = 
  \left(
    \begin{array}{cc}
      \cos^2\phi & \sin^2\phi\\
      \sin^2\phi & \cos^2\phi
    \end{array}
  \right)
\end{equation}
with eigenvalues 1 and $\cos^2\phi-\sin^2\phi$.  Thus the gap is equal to
$2\min(1-\sin^2\phi, \sin^2\phi)$ and its distribution is uniform
on the interval $[0,1]$ (which agrees with \Fref{fig:hist1}).

\subsection{Mean of the second largest eigenvalue}

\begin{theorem}
  Let $\lambda_2(\M)$ be the second largest (by modulus) eigenvalue of
  the random doubly stochastic matrix $\M$ drawn from $\mathrm{DS}(N)$
  with the probability induced by the correspondence $S:
  \mathrm{CUE}(N) \to \mathrm{DS}(N)$.  Then
  \begin{equation}
    \label{eq:theo1}
    \E|\lambda_2|\to0 \qquad\mbox{as } N\to\infty.
  \end{equation}
  As a consequence, the gap distribution $f_N(g)$ converges to
  $\delta(g-1)$ in the sense of distributions as $N\to\infty$.
\end{theorem}

\begin{proof}
  The eigenvalues of $\M\in\textrm{DS}(N)$ are, in general, complex.
  It would be more convenient to deal with nonnegative real
  eigenvalues.  The trick is to consider the matrix $\A = \M^{T}\M$
  which, due to the semigroup property is also doubly stochastic.  In
  addition, it is symmetric (all eigenvalues are real) and positive
  definite (i.e. all eigenvalues are nonnegative).  To relate the
  eigenvalues of the matrices $\A$ and $\M$ we write for the second
  largest eigenvalue of $\A$,
  \begin{equation}
    \label{eq:evalsAM}
    \lambda_2(\A) = \max_{|x|=1,\ (x,e)=0} (\A x, x) \geq (\A y, y) =
    (\M y, \M y) = |\lambda_2(\M)|^2,
  \end{equation}
  where $e = (1,\ldots, 1)^T$ is the eigenvector of both $\A$ and $\M$
  with the eigenvalue 1, and $y$ is the eigenvector of $\M$ corresponding
  to $\lambda_2(\M)$.
  From (\ref{eq:evalsAM}) and the inequality
  $(\E \xi)^2 < \E(\xi^2)$ we infer that to prove the Theorem it is
  enough to show that the mean of $|\lambda_2(\A)|$ converges to zero as
  $N\to\infty$.
  
  The simplest way to estimate the second largest eigenvalue of $\A$
  is to compute its trace.  Since $\A$ is positively defined, we have
  \begin{equation}
    \label{eq:traceAn}
    \Tr\A^n \geq 1 + \lambda_2(\A)^n.
  \end{equation}
  Thus if we can find such $n$ that $\E(\Tr\A^n)\to1$, it will imply
  that $\E(\lambda_2(\A))\to0$.  It turns out that it is enough to
  take $n=2$.  However let us consider $n=1$ too.
  \begin{equation}
    \label{eq:traceA1}
    \Tr\A = \sum_{i=1}^N (\M^T\M)_{i,i} = \sum_{i,j=1}^N
    (\M^T)_{i,j}(\M)_{j,i} = \sum_{i,j=1}^N
    M^2_{j,i} = \sum_{i,j=1}^N |U_{i,j}|^4.
  \end{equation}
  And calculating the mean
  \begin{equation}
    \label{eq:meantraceA1}
    \E(\Tr\A) = \sum_{i,j=1}^N \E|U_{i,j}|^4 = N^2 \E|U_{1,1}|^4,
  \end{equation}
  where due to the invariance of measure of CUE$(N)$ the means of the
  different matrix elements are equal.  To calculate $\E|U_{1,1}|^4$
  one can either integrate over the measure of CUE$(N)$, as was done
  in \cite{Ull}, or apply various invariance considerations
  \cite{Mel}.  The result is $\E|U_{1,1}|^4 = 2/(N^2+N)$ and therefore
  $\E(\Tr\A) \to 2$ as $N\to\infty$.  This agrees with the numerical
  observation that $\E|\lambda_2(\M)| \propto 1/\sqrt{N}$: the
  eigenvalues of $\A$ are then of order $1/N$ and there are $N-1$ of
  them (not counting the 1).

  In the case $n=2$ we have
  \begin{eqnarray}
    \label{eq:traceA2}
    \nonumber
    \fl\Tr\A^2 = \sum_{i,j,k,l=1}^N
    (\M^T)_{i,j}(\M)_{j,k}(\M^T)_{k,l}(\M)_{l,m}
    = \sum_{i,j,k,l=1}^N
    |U_{j,i}U_{j,k}U_{l,i}U_{l,k}|^2\\
    \nonumber
    \lo{=} \sum_{i\neq k,j\neq l}
    |U_{j,i}U_{j,k}U_{l,i}U_{l,k}|^2  
    + \sum_{i=k,j\neq l} |U_{j,i}U_{l,i}|^4 \\
    + \sum_{i\neq k,j=l} |U_{j,i}U_{j,k}|^4
    + \sum_{i=k,j=l} |U_{j,i}|^8.
  \end{eqnarray}
  Applying the averaging we again find that due to the invariance of
  the measure all contributions from the first sum are the same and
  there are $(N^2-N)^2$ of them; there are $2N^2(N-1)$ contributions
  from the second and the third sum (the contributions are equal) and
  $N^2$ contributions from the last.  Counting the number of terms in
  each sum, we write
  \begin{eqnarray}
    \label{eq:meantraceA2}
    \nonumber
    \fl\E\Tr\A^2 = (N^2-N)^2 \E|U_{1,1}U_{1,2}U_{2,1}U_{2,2}|^2\\
    + 2N^2(N-1) \E|U_{1,1}U_{1,2}|^4
    + N^2 \E|U_{1,1}|^8,
  \end{eqnarray}
  where the averages can be calculated using \cite{Mel},
  \begin{eqnarray}
    \label{eq:averages}
    \E|U_{1,1}U_{1,2}U_{2,1}U_{2,2}|^2 
    &=& \frac{N^2+N+2}{N^2(N^2-1)(N+2)(N+3)},\\
    \E|U_{1,1}U_{1,2}|^4 &=& \frac{4}{N(N+1)(N+2)(N+3)},\\
    \E|U_{1,1}|^8 &=& \frac{24}{N(N+1)(N+2)(N+3)}.
  \end{eqnarray}
  Bringing everything together, we obtain
  \begin{equation}
    \label{eq:result_trace}
    \E\Tr\A^2 = \frac{N^3+8N^2+17N-2}{(N+1)(N+2)(N+3)} = 1 +
    \Or\brr{\frac1N}, 
  \end{equation}
  which effectively finishes the proof.
\end{proof}

\section{Shape of the distribution of $|\lambda_2(\M)|$}

It is now clear that the distributions $f_N(|\lambda_2|)$ have
singular limit.  The natural question to ask, then, is what is the
shape of $f_N(|\lambda_2|)$ for finite, but large, $N$ and whether
there are sequences of normalizing coefficients $a_N$ and $b_N$ such
that the random variables $(|\lambda_2|-a_N)/b_N$ have continuous
limiting distribution.  While we are unable to give a definite answer,
we can make a guess.  Since
$|\lambda_2|=\max_{j=2,\ldots,N}|\lambda_j|$, it is not unnatural to
conjecture that for large $N$ the distribution of the second largest
eigenvalue $|\lambda_2|$ is well approximated by the Generalized
Extreme Value distribution (GEV) \cite{LeaLR,Col}, specified by its
cumulative distribution function
\begin{equation}
  \label{eq:gev_def}
  G(x) = \exp\brf{-\left[1+\xi\brr{\frac{x-a}{b}}\right]^{-1/\xi}},
\end{equation}
for some ($N$-dependent) $\xi$, $a$ and $b$: the GEV distribution is
used to model maxima of sequences of random variables.

To check our conjecture we fit the numerical data for $N=80$ to the
form~(\ref{eq:gev_def}) using an {\tt Splus} routine by
Coles \cite{Col}.  The routine minimizes the log-likelihood function
\begin{equation}
  \label{eq:gev_lik}
  l(a,b,\xi) = \sum_{i=1}^k 
  \brf{-\ln b\left[1+\xi\brr{\frac{x_i-a}{b}}\right]^{1+1/\xi} -
    \left[1+\xi\brr{\frac{x-a}{b}}\right]^{-1/\xi}},
\end{equation}
where $x_i$ is the sequence of observations of $|\lambda_2|$.  The
result of the routine is plotted as the probability density function
$G'(x)$ for fitted $\xi=-0.07$, $a$ and $b$ against the histogram of
the numerical data for $|\lambda_2|$.  One can see that the agreement
is very good.  The probability and quantile plots, which we do not
present here, also show good agreement.  As $N$ is taken to infinity,
the values of $a$ and $b$ decay to zero, so that the limit of the
distribution is the step function, and $\xi$ saturates at a nonzero
value.

\section{Conclusions}

We have shown that if we take element-wise modulus squared of a large
unitary matrix $\U$, the resulting doubly stochastic matrix $\M$ will
have large spectral gap: with high probability all $N-1$ ``free''
eigenvalues will have small modulus.  In particular, it means that the
Markov chain corresponding to $\M$ will quickly relax from any
initial distribution into the uniform stationary state.

While our observation is by no means a proof of Tanner's \cite{Tan_p}
conjecture that the families of graphs with large gap in their
``classical'' spectrum will adhere to RMT predictions, it is another
powerful argument in its favour.  We also note that the unitary
evolution matrix of a general graph is sparse due to the topological
restrictions \cite{KotS1,KotS2}.  Thus, although arbitrary $\U$ can be
considered as the scattering matrix for an $N$-star graph, the
generalisation of the gap distribution to graphs with less trivial
topology still remains to be studied.

The results for the mean of the traces of $\M^T\M$ and $(\M^T\M)^2$,
obtained in the course of our proof, completely agree with the
numerical observation that the mean gap increases with $N$ like
$1-\Or\brr{N^{-1/2}}$.  We also presented numerical evidence in favour
of the claim that for finite $N$ the distribution of the second
largest eigenvalue is well approximated by the generalized extreme
value distribution.


\ack The author is very grateful to M.~Grinfeld, B.~Gutkin,
J.P.~Keating, U.~Smilansky, N.~Snaith, M.G.~Stepanov, S.~Subba~Rao~Tata,
G.~Tanner and K.~\.Zyczkowski.  Without their suggestions and
encouragement this work would not be possible.  The author was supported
by the Israel Science Foundation, a Minerva grant, and the Minerva Center
for Nonlinear Physics.

\newpage

\begin{figure}[b]
  \includegraphics[scale=0.7,angle=270]{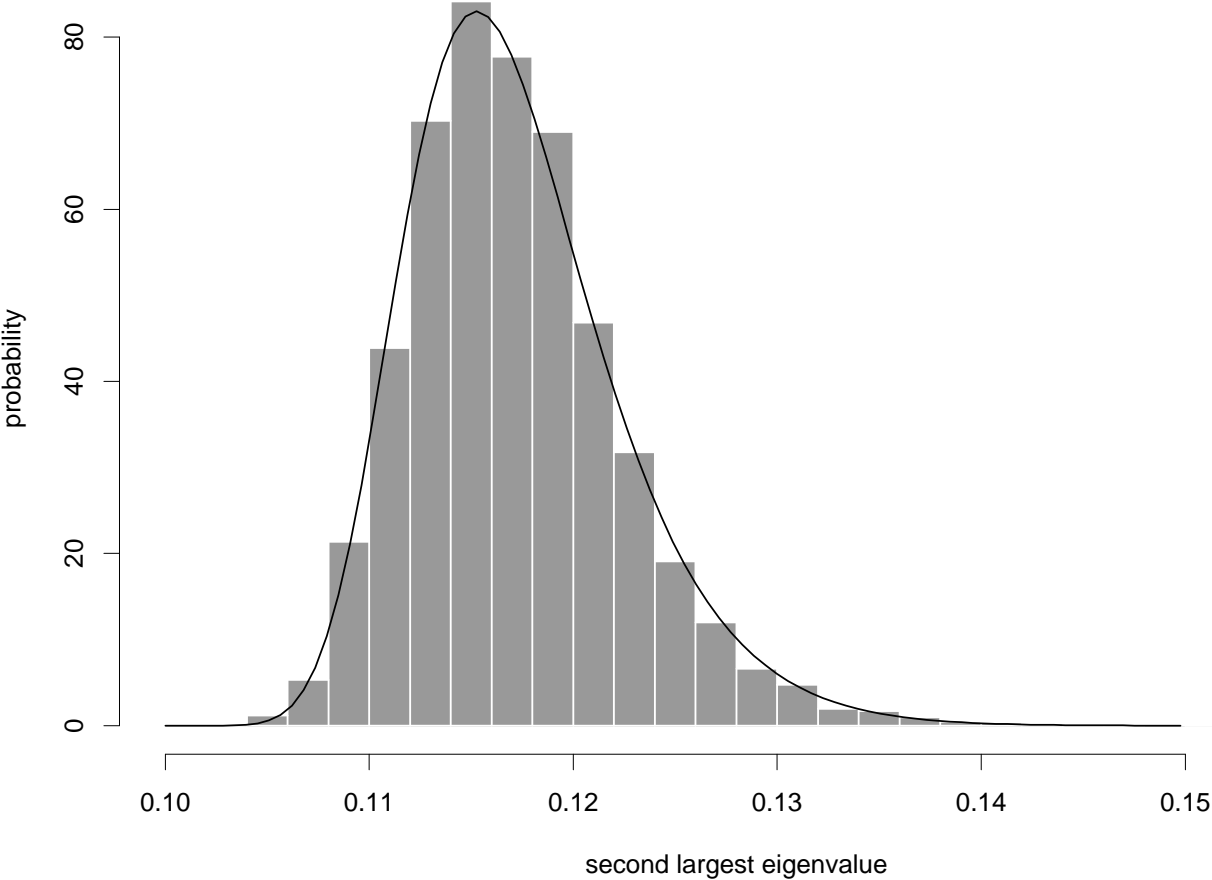} 
  \caption{Histogram of the numerical data from the simulation of
    $|\lambda_2|$ for $N=80$ and the fitted GEV probability density
    function.}
  \label{fig:gev}
\end{figure}

\end{document}